\documentclass[preprint,5p,twocolumn,number,11pt]{elsarticle}

\usepackage{graphics}

\usepackage{amssymb}
\usepackage{amsthm}

\usepackage{lineno}
\usepackage{ulem}
\usepackage{tabularx}
\usepackage{textcomp}
\usepackage[english]{babel}
\usepackage{microtype}

\journal{Astroparticle Physics}

\begin{document}

\begin{frontmatter}


\title{Improved absolute calibration of LOPES measurements and its impact on the comparison with REAS 3.11 and CoREAS simulations}

\author[a]{W.D.~Apel}
\author[b]{J.C.~Arteaga-Vel\'azquez}
\author[c]{L.~B\"ahren}
\author[a]{K.~Bekk}
\author[d]{M.~Bertaina}
\author[e]{P.L.~Biermann}
\author[a,f]{J.~Bl\"umer}
\author[a]{H.~Bozdog}
\author[g]{I.M.~Brancus}
\author[d,h]{E.~Cantoni}
\author[d]{A.~Chiavassa}
\author[a]{K.~Daumiller}
\author[i]{V.~de~Souza}
\author[d]{F.~Di~Pierro}
\author[a]{P.~Doll}
\author[a]{R.~Engel}
\author[c,e,j]{H.~Falcke}
\author[f]{B.~Fuchs}
\author[l]{H.~Gemmeke}
\author[m]{C.~Grupen}
\author[a]{A.~Haungs}
\author[a]{D.~Heck}
\author[a]{R.~Hiller}
\author[j]{J.R.~H\"orandel}
\author[e]{A.~Horneffer}
\author[f]{D.~Huber}
\author[a]{T.~Huege}
\author[n]{P.G.~Isar}
\author[k]{K.-H.~Kampert}
\author[f]{D.~Kang}
\author[l]{O.~Kr\"omer}
\author[j]{J.~Kuijpers}
\author[f]{K.~Link}
\author[o]{P.~{\L}uczak}
\author[f]{M.~Ludwig}
\author[a]{H.J.~Mathes}
\author[f]{M.~Melissas}
\author[h]{C.~Morello}
\author[p]{S.~Nehls}
\author[a]{J.~Oehlschl\"ager}
\author[f]{N.~Palmieri}
\author[a]{T.~Pierog}
\author[k]{J.~Rautenberg}
\author[a]{H.~Rebel}
\author[a]{M.~Roth}
\author[l]{C.~R\"uhle}
\author[g]{A.~Saftoiu}
\author[a]{H.~Schieler}
\author[l]{A.~Schmidt}
\author[a]{S.~Schoo}
\author[a]{F.G.~Schr\"oder\corref{cor}}
\author[q]{O.~Sima}
\author[g]{G.~Toma}
\author[h]{G.C.~Trinchero}
\author[a]{A.~Weindl}
\author[a]{J.~Wochele}
\author[o]{J.~Zabierowski}
\author[e]{J.A.~Zensus}

\address[a]{Institut f\"ur Kernphysik, Karlsruhe Institute of Technology (KIT), Karlsruhe, Germany}
\address[b]{Instituto de F\'isica y Matem\'aticas, Universidad Michoacana, Morelia, Michoac\'an, Mexico}
\address[c]{ASTRON, Dwingeloo, The Netherlands}
\address[d]{Dipartimento di Fisica, Universit\`a degli Studi di Torino, Torino, Italy}
\address[e]{Max-Planck-Institut f\"ur Radioastronomie, Bonn, Germany}
\address[f]{Institut f\"ur Experimentelle Kernphysik, Karlsruhe Institute of Technology (KIT), Karlsruhe, Germany}
\address[g]{National Institute of Physics and Nuclear Engineering, Bucharest-Magurele, Romania}
\address[h]{Osservatorio Astrofisico di Torino, INAF Torino, Torino, Italy}
\address[i]{Instituto de F\'isica de S\~ao Carlos, Universidade de S\~ao Paulo, S\~ao Carlos, Brasil}
\address[j]{Department of Astrophysics, Radboud University Nijmegen, AJ Nijmegen, The Netherlands}
\address[k]{Fachbereich C, Physik, Bergische Universit\"at Wuppertal, Wuppertal, Germany}
\address[l]{Institut f\"ur Prozessdatenverarbeitung und Elektronik, Karlsruhe Institute of Technology (KIT), Karlsruhe, Germany}
\address[m]{Faculty of Natural Sciences and Engineering, Universit\"at Siegen, Siegen, Germany}
\address[n]{Institute for Space Sciences, Bucharest-Magurele, Romania}
\address[o]{Department of Astrophysics, National Centre for Nuclear Research, {\L}\'{o}d\'{z}, Poland}
\address[p]{Studsvik Scandpower GmbH, Hamburg, Germany}
\address[q]{Department of Physics, University of Bucharest, Bucharest, Romania}

\cortext[cor]{Corresponding author: frank.schroeder@kit.edu}

\begin{abstract}
LOPES was a digital antenna array detecting the radio emission of cosmic-ray air showers. The calibration of the absolute amplitude scale of the measurements was done using an external, commercial reference source, which emits a frequency comb with defined amplitudes. Recently, we obtained improved reference values by the manufacturer of the reference source, which significantly changed the absolute calibration of LOPES. We reanalyzed previously published LOPES measurements, studying the impact of the changed calibration. The main effect is an overall decrease of the LOPES amplitude scale by a factor of $2.6 \pm 0.2$, affecting all previously published values for measurements of the electric-field strength. This results in a major change in the conclusion of the paper \lq Comparing LOPES measurements of air-shower radio emission with REAS 3.11 and CoREAS simulations\rq~published in Astroparticle Physics 50-52 (2013) 76-91 \cite{LOPESlateralComparison2013}: 
With the revised calibration, LOPES measurements now are compatible with CoREAS simulations, but in tension with REAS 3.11 simulations. Since CoREAS is the latest version of the simulation code incorporating the current state of knowledge on the radio emission of air showers, this new result indicates that the absolute amplitude prediction of current simulations now is in agreement with experimental data.
\end{abstract}

\begin{keyword}
cosmic rays \sep extensive air showers \sep radio emission \sep LOPES \sep absolute calibration
\end{keyword}

\end{frontmatter}


\section{Improved calibration of LOPES}
LOPES was the radio extension of the KASCADE-Grande particle-detector array for cosmic-ray air showers \cite{ApelArteagaBadea2010, FalckeNature2005}. Triggered by KASCADE-Grande, it detected the radio emission of the same air showers as measured by the particle-detector array in the effective frequency band of $43-74\,$MHz. 

For calibration of LOPES, we used an externally calibrated reference source consisting of a signal generator and a biconical antenna \cite{NehlsHakenjosArts2007}. This reference source emits a train of equidistant pulses, which in the frequency domain corresponds to a comb with $1\,$MHz spacing. The manufacturer provided reference values for this source with an overall uncertainty of $2.5\,$dB, which was the main contribution to the total two-sigma uncertainty of roughly $35\,\%$ for the LOPES amplitude scale as published earlier \cite{NehlsHakenjosArts2007, LOPESlateralComparison2013}. We provided the reference source also to other experiments, namely LOFAR \cite{SchellartLOFAR2013} and Tunka-Rex \cite{TunkaRexNIM2015}, to have a consistent absolute amplitude scale between these experiments. Their results can now be compared on an absolute level, which was a problem for historic experiments \cite{atrashkevich}. 

In this context, the reference source has been re-measured by the manufacturer \cite{calibrationStandard}. The old calibration values used for previous LOPES publications characterized the reference source for free-field conditions. This means that a reflective ground in a horizontal setup was used in the manufacturer's calibration measurement of the reference source. Such a setup is useful for ground-based communication applications, but leads to significant interference effects. 
The manufacturer's measurement was performed at several heights above ground, finally taking the maximum value. Consequently, the effect of constructive interference was significantly enhanced. Because ground effects are already taken into account in the simulation of the LOPES antennas used for the evaluation of air-shower measurements, this led to a significant overestimation of the amplitudes measured with LOPES. To first approximation, a factor of two difference is expected between free-field conditions with constructive interference of ground reflections and the now-used free-space conditions corresponding to no reflections.

Based on a new measurement of our reference source (not just of a source of the same type), the new calibration values have now been determined for free-space conditions, which better match the situation of air showers. The two-sigma scale uncertainty of the amplitude is still given as $2.5\,$dB by the manufacturer. This corresponds to a one-sigma uncertainty of $16\,\%$ for the amplitude (field strength) scale. This uncertainty covers potential repeated measurements under equal conditions, not the change between different conditions. At the more sensitive instrument LOFAR it has been checked that these new free-space reference values are consistent within the scale uncertainty with an independent calibration on galactic background \cite{NellesLOFARcalibration2015}. 

It turns out that the new reference values lead to a significant change of the LOPES amplitude scale.

\section{Impact on shower reconstruction}
We have analyzed the impact of the improved calibration on the reconstruction of the radio emission measured with LOPES for a data set of about $500$ events recorded with east-west aligned antennas which has been used in reference \cite{LOPESlateralComparison2013}. The number of selected events is slightly lower, since due to the improved calibration a few events close to threshold do not pass anymore the quality cuts. Moreover, we have made a few smaller improvements in the analysis pipeline \cite{LinkICRC2015, SchroederICRC2015} which, however, have no significant impact on the results reported here.

\begin{figure}[t]
  \centering
    \includegraphics[width=0.99\columnwidth]{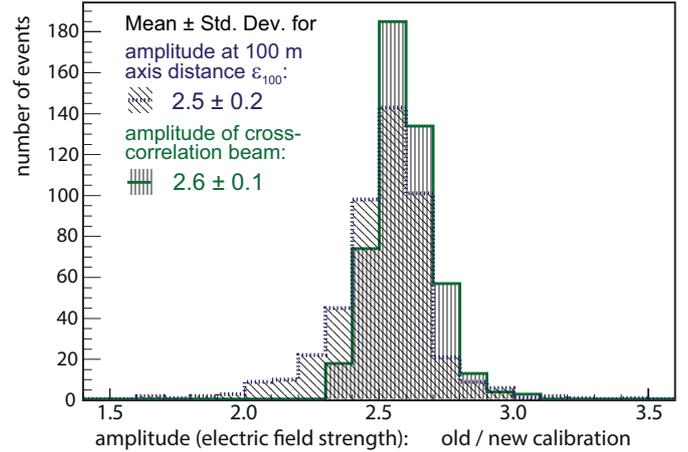}
  \caption{Change in amplitude (electric-field strength) of LOPES measurements due to the new, improved absolute calibration for the height of the interferometric cross-correlation beam and $\epsilon_{100}$. The change is not exactly equal for all events, because the calibration is slightly frequency-dependent, and each LOPES measurement might have a different frequency spectrum, since this depends on the shower geometry.}
   \label{fig_amplitudeChange}
\end{figure}

\begin{figure*}[t]
  \centering
    \includegraphics[width=0.4\textwidth]{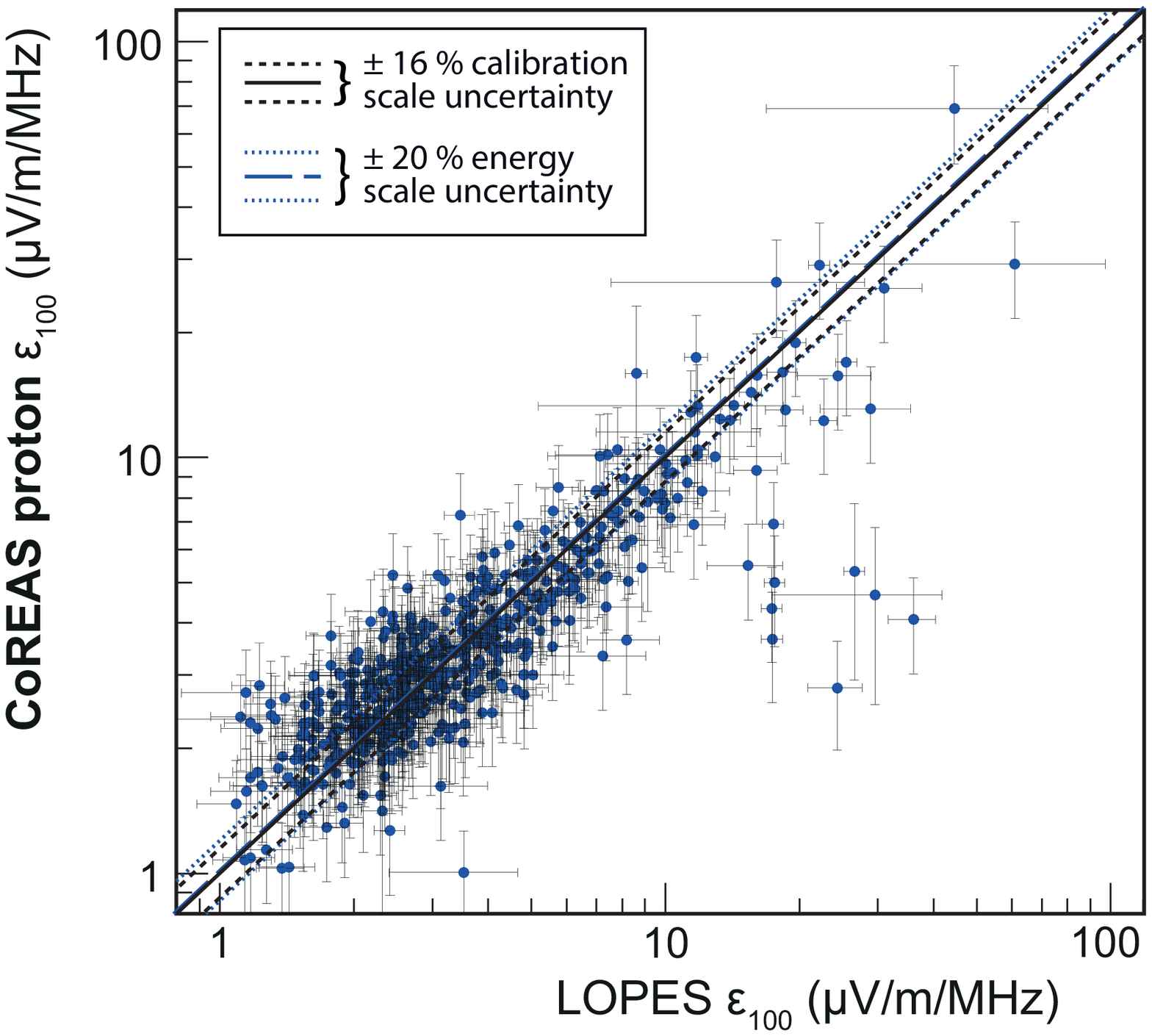}
    \hfill
    \includegraphics[width=0.49\textwidth]{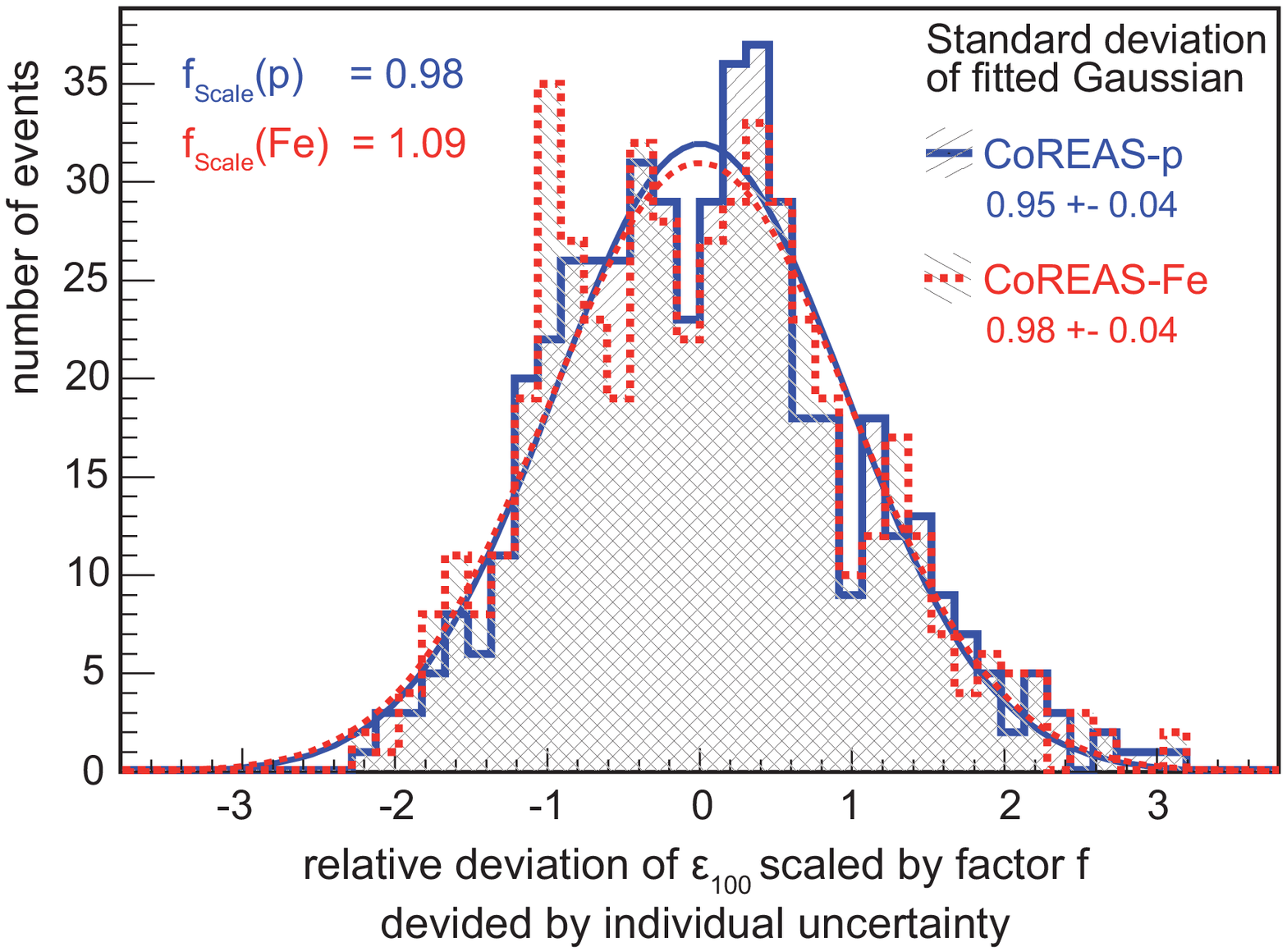}
  \caption{Comparison of the amplitude at $100\,$m, $\epsilon_{100}$, between LOPES measurements and CoREAS simulations. Left: scatter plot of for all $502$ LOPES events compared to CoREAS simulations with proton primaries. The result for iron nuclei as primaries is very similar. Right: Deviation of each event divided by the uncertainty of the event, after multiplying the simulated amplitudes with a scaling factor $f$ ($0.98$ for proton, and $1.09$ for iron simulations) such that the distributions are centered around 0. A Gaussian fitted to the histogram has a width of approximately $1$ ($0.95 \pm 0.04$ for proton, and $0.98 \pm 0.04$ for iron simulations), which means that the spread visible in the left figure corresponds to the spread expected by the measurement uncertainties, except of the few outliers of unknown origin.}
   \label{fig_epsilon100comparison}
\end{figure*}

For the selection of events we apply the digital interferometric technique of cross-correlation beamforming, and then use pulse measurements in individual antennas for further analyses. As in references \cite{ApelArteagaAsch2010, ApelLOPES_MTD2012, LOPESlateralComparison2013} we fit measured lateral distributions of amplitude versus distance to shower axis $d$ with an exponential function $\epsilon(d) = \epsilon_{100}\,\exp[-\eta(d - 100\,\mathrm{m})]$. This function has two parameters: $\epsilon_{100}$, the amplitude (electric-field strength) at $100\,$m distance from the shower axis, and $\eta$ describing the slope of the lateral distribution. 

Fig.~\ref{fig_amplitudeChange} shows that the amplitude scale of both the cross-correlation beam and of $\epsilon_{100}$ is lowered by the average factor of $2.6 \pm 0.2$ due to the improved calibration (combining both values in Fig.~\ref{fig_amplitudeChange}). The shape of the lateral distribution remains practically unchanged: the average value for $\eta$ changes by less than its systematic uncertainty of about $1\,$km\textsuperscript{-1} (not shown here). Consequently, previously published LOPES results remain valid, but all field-strength values have to be divided by $2.6 \pm 0.2$, which affects, e.g., the proportionality factor in published formulas for energy reconstruction \cite{ApelLOPES_LDF_Xmax2014, HornefferICRC2007, SchroederLOPES_ARENA2012}.

\section{Significance for comparison to simulations}
The lowered amplitude scale has important consequences for the comparison of simulated and measured lateral distributions. In reference \cite{LOPESlateralComparison2013}, we compared LOPES data to the now obsolete predictions by REAS 3.11 \cite{LudwigHuege2010} and those of its state-of-the-art successor CoREAS \cite{HuegeCoREAS_ARENA2012}, which predicts roughly two times lower amplitudes $\epsilon_{100}$ than REAS 3.11. Both, REAS 3.11 and CoREAS, are microscopic simulation codes based on CORSIKA \cite{HeckKnappCapdevielle1998}, and implicitly include all emission mechanisms known to be relevant. The main difference is that CoREAS calculates the radio emission directly during the simulation of the particle shower, while REAS 3.11 calculates it afterwards based on histograms which neglect some correlations of the particles in the shower development.

\begin{figure}
  \centering
    \includegraphics[width=0.4\textwidth]{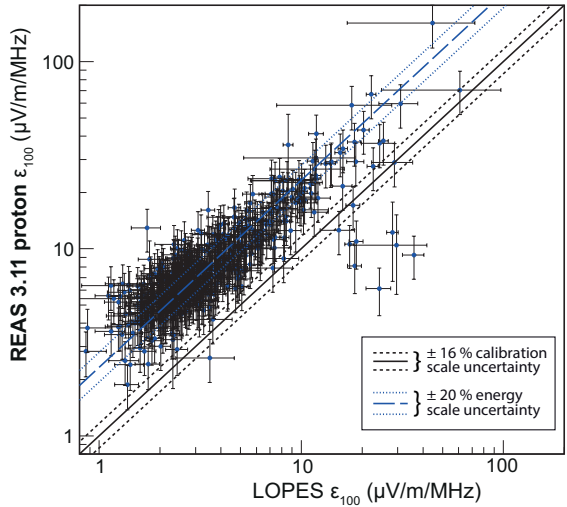}
  \caption{Comparison of $\epsilon_{100}$ between LOPES measurements and REAS 3.11 simulations.}
   \label{fig_amplitudeChangeREAS}
\end{figure}

With the improved calibration the amplitude parameter $\epsilon_{100}$ of the CoREAS simulations almost perfectly matches the measured data (event-by-event comparison and histograms in Fig.~\ref{fig_epsilon100comparison}). The mean deviation is only $2\,\%$ with protons as primary particles, and $9\,\%$ for iron nuclei as primary particles. Both deviations are much smaller than the remaining one-sigma uncertainty of the LOPES amplitude scale of approximately $16\,\%$.

For REAS 3.11 the situation is different (event-by-event comparison in Fig.~\ref{fig_amplitudeChangeREAS}, histograms not shown). The scaling factor needed to bring REAS 3.11 simulations in perfect agreement with LOPES measurements now is $f = 0.43$ and $f = 0.46$ for proton and iron primaries, respectively. Even including an additional $20\,\%$ uncertainty for the KASCADE-Grande energy scale \cite{ApelGrandeEnergySpectrum2012} used as input for the simulations, this is at tension with the measurements.

In other aspects tested in reference \cite{LOPESlateralComparison2013}, in particular with respect to the slope parameter $\eta$ and its dependence on the geometry, LOPES measurements continue to be compatible with the simulations. However, for the dependence of the amplitude on the zenith angle, we observe a slight difference between measurements and CoREAS simulations reported in reference \cite{ApelLOPES_LDF_Xmax2014}. This difference remains, i.e.~the scaling factor $f$ varies by several $10\,\%$ over zenith angle \cite{LinkICRC2015, SchroederICRC2015}. Nevertheless, this is not necessarily a problem of the simulations, since the effect could be due to systematic uncertainties of the antenna model used for conversion of LOPES measurements. Hence, this should be checked with other experiments using different antennas. Consequently, all tested aspects of CoREAS are compatible with LOPES measurements, which is remarkable because CoREAS does not feature any free parameters tuned against the data.

\small{
\section*{Acknowledgments}
LOPES and KASCADE-Grande have been supported by the German Federal Ministry of Education and Research. KASCADE-Grande is partly supported by the MIUR and INAF of Italy, the Polish Ministry of Science and Higher Education and by the Romanian Authority for Scientific Research UEFISCDI (PNII-IDEI grant 271/2011). The present study is supported by grant VH-NG-413 of the Helmholtz association and by the ’Helmholtz Alliance for Astroparticle Physics - HAP’ funded by the Initiative and Networking Fund of the Helmholtz Association, Germany. 
}


\begin{thebibliography}{00}


\bibitem{LOPESlateralComparison2013}
\bibinfo{author}{{Apel}, W.~D.},
\bibinfo{author}{{et al.~- LOPES Collaboration}},
\newblock \bibinfo{journal}{Astropart. Phys.} \bibinfo{volume}{50-52}
 (\bibinfo{year}{2013}) \bibinfo{pages}{76-91}.

\bibitem{ApelArteagaBadea2010}
\bibinfo{author}{{Apel}, W.~D.},
\bibinfo{author}{{et al.~- KASCADE-Grande Collaboration}},
\newblock \bibinfo{journal}{Nucl. Instr. Meth. A} \bibinfo{volume}{620}
  (\bibinfo{year}{2010{\natexlab{a}}}) \bibinfo{pages}{202-216}.  

  
\bibitem{FalckeNature2005}  
\bibinfo{author}{{Falcke}, H.},
\bibinfo{author}{et al.~- LOPES Collaboration},
\newblock \bibinfo{journal}{Nature} \bibinfo{volume}{435}
  (\bibinfo{year}{2005}) \bibinfo{pages}{313-316}.
  
\bibitem{NehlsHakenjosArts2007}
\bibinfo{author}{{Nehls}, S.},
\bibinfo{author}{et al.},
\newblock \bibinfo{journal}{Nucl. Instr. Meth. A} \bibinfo{volume}{589}
  (\bibinfo{year}{2008}) \bibinfo{pages}{350-361}.
  
\bibitem{SchellartLOFAR2013}  
\bibinfo{author}{{Schellart}, P.},
\bibinfo{author}{et al.~- LOFAR Collaboration},
\newblock \bibinfo{journal}{Astronomy and Astrophysics} \bibinfo{volume}{560}
  (\bibinfo{year}{2013}) \bibinfo{pages}{A98}.

  
\bibitem{TunkaRexNIM2015}
\bibinfo{author}{{Bezyazeekov}, P.~A.},
\bibinfo{author}{{et al.~- Tunka-Rex Collaboration}},
\newblock \bibinfo{journal}{Nucl. Instr. Meth. A} \bibinfo{volume}{802}
  (\bibinfo{year}{2015}) \bibinfo{pages}{89-96}.
  
\bibitem{atrashkevich}
\bibinfo{author}{{Atrashkevich}, V.~B.},
\bibinfo{author}{{et al.}},
\newblock \bibinfo{journal}{Sov. J. Nucl. Phys.} \bibinfo{volume}{28}
  (\bibinfo{year}{1978}) \bibinfo{pages}{366}.
  
    
\bibitem{calibrationStandard}
The calibration is done in line with EN ISO/IEC 17025.
  
 \bibitem{NellesLOFARcalibration2015}  
\bibinfo{author}{{Nelles}, A.},
\bibinfo{author}{et al.~- LOFAR Collaboration},
\newblock \bibinfo{journal}{JINST} \bibinfo{volume}{10}
  (\bibinfo{year}{2015}) \bibinfo{pages}{P11005}. 
  
\bibitem{LinkICRC2015}
\bibinfo{author}{{Link}, K.},
\bibinfo{author}{{et al.~- LOPES Collaboration}},
\newblock \bibinfo{journal}{Proceeding of Science} \bibinfo{volume}{ICRC2015}
  (\bibinfo{year}{2015}) \bibinfo{pages}{311}. 
  
\bibitem{SchroederICRC2015}
\bibinfo{author}{{Schr\"oder}, F.G.},
\bibinfo{author}{{et al.~- LOPES Collaboration}},
\newblock \bibinfo{journal}{Proceeding of Science} \bibinfo{volume}{ICRC2015}
  (\bibinfo{year}{2015}) \bibinfo{pages}{317}.
 

\bibitem{ApelArteagaAsch2010}
\bibinfo{author}{{Apel}, W.~D.},
\bibinfo{author}{{et al.~- LOPES Collaboration}},
\newblock \bibinfo{journal}{Astropart. Phys.} \bibinfo{volume}{32}
 (\bibinfo{year}{2010}) \bibinfo{pages}{294-303}.
 
\bibitem{ApelLOPES_MTD2012}
\bibinfo{author}{{Apel}, W.~D.},
\bibinfo{author}{{et al.~- LOPES Collaboration}},
\newblock \bibinfo{journal}{Phys. Rev. D}  \bibinfo{volume}{85}
  (\bibinfo{year}{2012}) \bibinfo{pages}{071101(R)}. 

\bibitem{ApelLOPES_LDF_Xmax2014}
\bibinfo{author}{{Apel}, W.~D.},
\bibinfo{author}{{et al.~- LOPES Collaboration}},
\newblock \bibinfo{journal}{Phys. Rev. D}  \bibinfo{volume}{90}
  (\bibinfo{year}{2014}) \bibinfo{pages}{062001}. 
  
\bibitem{HornefferICRC2007}
\bibinfo{author}{{Horneffer}, A.},
\bibinfo{author}{{et al.~- LOPES Collaboration}},
\newblock in: \bibinfo{booktitle}{Proc. of the 30th International Cosmic Ray Conference 2007, Merida, Mexico},
  volume~\bibinfo{volume}{4}, pp. \bibinfo{pages}{83-86}.
  
  
\bibitem{SchroederLOPES_ARENA2012}
\bibinfo{author}{{Schr\"oder}, F.~G.},
\bibinfo{author}{{et al.~- LOPES Collaboration}},
\newblock \bibinfo{journal}{AIP Conference Proceeding}   \bibinfo{volume}{1535} (\bibinfo{year}{2013}) \bibinfo{pages}{78-83}.


\bibitem{LudwigHuege2010}
\bibinfo{author}{{Ludwig}, M., {Huege}, T.},
\newblock \bibinfo{journal}{Astropart. Phys.} \bibinfo{volume}{34}
  (\bibinfo{year}{2011}) \bibinfo{pages}{438-446}.

  
\bibitem{HuegeCoREAS_ARENA2012}
\bibinfo{author}{{Huege}, T.},
\bibinfo{author}{{Ludwig}, M.},
\bibinfo{author}{{James}, C.},
\newblock \bibinfo{journal}{AIP Conference Proceeding}  \bibinfo{volume}{1535} (\bibinfo{year}{2013}) \bibinfo{pages}{128-132}.


\bibitem{HeckKnappCapdevielle1998}
\bibinfo{author}{{Heck}, D. et~al.}.
\newblock FZKA Report 6019, Forschungszentrum Karlsruhe (1998). 

\bibitem{ApelGrandeEnergySpectrum2012}
\bibinfo{author}{{Apel}, W.~D.},
\bibinfo{author}{{et al.~- KASCADE-Grande Collaboration}},
\newblock \bibinfo{journal}{Astropart. Phys.} \bibinfo{volume}{36}
  (\bibinfo{year}{2012}) \bibinfo{pages}{183-194}.  

  


  




\end{thebibliography}
\end{document}